\documentclass[epjCONF, onecolumn]{svjour}
\usepackage{graphics}
\usepackage[varg]{txfonts} % Times fonts
\usepackage[latin1]{inputenc}
\session-title{Conference Title, to be filled}
\begin{document}
\title{The current population of benchmark brown dwarfs}
\author{A.C. Day-Jones\inst{^1,^2}\fnmsep\thanks{\email{adjones@das.uchile.cl}} \and D.J. Pinfield\inst{^2} \and M.T. Ruiz\inst{^1} \and B. Burningham\inst{^2} \and Z.H. Zhang\inst{^2} \and H.R.A. Jones\inst{^2}\and M.C. G\'alvez-Ortiz \inst{^2} \and J. Gallardo\inst{^1} \and J.R.A. Clarke\inst{^2} \and J.S. Jenkins\inst{^1} }
\institute{Departamento de Astronomia, Universidad de Chile, Santiago, Chile \and Centre for Astrophysics Research, University of Hertfordshire, Hatfield, AL10 9AB, UK}
\abstract{ The number of brown dwarfs (BDs) now identified tops
  700. Yet our understanding of these cool objects is still lacking,
  and models are struggling to accurately reproduce observations. What
  is needed is a method of calibrating the models, BDs whose
  properties (e.g. age, mass, distance, metallicity) that can be
  independently determined can provide such calibration. The ability
  to calculate properties based on observables is set to be of vital
  importance if we are to be able to measure the properties of
  fainter, more distant populations of BDs that near-future surveys
  will reveal, for which ground based spectroscopic studies will
  become increasingly difficult. We present here the state of the
  current population of age benchmark brown dwarfs.  }
%end of abstract
%
\maketitle

\section{Benchmark brown dwarfs}
\label{benchmarks}
The BD population in the galactic neighborhood provides a range of environments and ages that could potentially be probed for those that are likely to be benchmark objects. Young clusters and moving groups, such as the Hyades, Pleiades and Praesepe and moving groups currently provide the wealth of identified benchmarks, e.g.\cite{bouvier08},\cite{rebolo95},\cite{magazzu98}. They can provide highly accurate ages, distances and metallicities that can measured from the cluster members. However they are only useful for probing the young population of $<$1~Gyr, after which disk heating mechanisms \cite{desimone04} cause cluster members to be dispersed into the disk. On rare occasions it may be possible to calculate the age of an isolated BD, from their kinematics (providing a distance and radial velocity are known), for example 2MASS0320-044 \cite{blake08}. The accuracy of ages measured this way however are not particularly high. The ages of a handful of other BDs have been measured using the lithium test \cite{magazzu93}, such as DENISp-J1228.2-1547 \cite{tinney97}, SDSSJ0423-0414 \cite{burgasser05},2M0850 \cite{kirkpatrick99} and Kelu1 \cite{ruiz97}.

Another source of benchmark BDs may come from close BD + BD binary pairs where the dynamical mass of the components can be calculated. However they can not, in general provide accurate age estimations, unless on the rare occasion they are found to be a member of a cluster or components of a multiple system (where the primary star has a known age), as is the case for HD 130948BC \cite{dupuy09}. A more abundant source of benchmark in binaries are where the host star can provide age, distance and in some cases metallicity. How useful these systems are however is dependent on how well we understand the nature and physics of the host stars.  Main-sequence (MS) star primaries are by far the most abundant binary hosts for BDs after BD + BD pairs, however their ages can be largely uncertain due to the convergence of evolutionary models on the MS e.g. \cite{girardi00},\cite{yi01}. Binary hosts with better age constraints are likely to be gained from more evolved stars, e.g. subgiants, giants and white dwarfs \cite{pinfield06}.  Whose physics are relatively well understood and their ages can be calculted from robust evolutionary models. These types of binary system are sensitive to ages covering that of the disk (1-10Gyrs, see \cite{pinfield06}), filling the age space that BDs in clusters can not probe.

%Your text comes here. Separate text sections with
\section{The current benchmark population}
\label{benchpop}

In order for a benchmark to be useful for models it must be able to be studied in detail, such that it should be possible to obtain a good (high S/N) spectrum of the individual BD. Some potentially useful BDs that are members of unresolved binaries, or those that may have undergone interaction with their host star, such as WD0137 \cite{maxted06},\cite{burleigh06} are not ideal for this purpose and we do not include them in our selection of benchmark BDs. We present in Fig~1. the BDs that have independently derived ages, with the parameters described by their various authors. We do not include very young cluster BDs, such as those in Orion, Taurus, IC328, $\rho$Ophiucus, Chameleon or Upper Sco, as the formation of BDs is not, as yet fully understood and models of such youthful BDs remain very unreliable. The accuracy of the measurements of these properties (particularly age) is clearly not alway high. What is needed now is objects with smaller associated uncertainties which will enable the accurate study of the relationship between BD observable characteristics and their physical properties. This is becoming more important as we discover fainter, more distant populations of BDs that will be increasingly difficult to follow-up with good spectroscopic measurements using current ground based instruments.

%\section{Section title}
%\label{sec:1}
%and \cite{RefJ}
%\subsection{Subsection title}
%\label{sec:2}
%as required. Don't forget to give each section
%and subsection a unique label (see Sect.~\ref{sec:1}).
%

\begin{figure}
%\resizebox{0.75\columnwidth}{!}{%
%\includegraphics{benchmarksbdspostercnecan.eps} }
\vspace{4.3cm}
\hspace{5.0cm}
\includegraphics{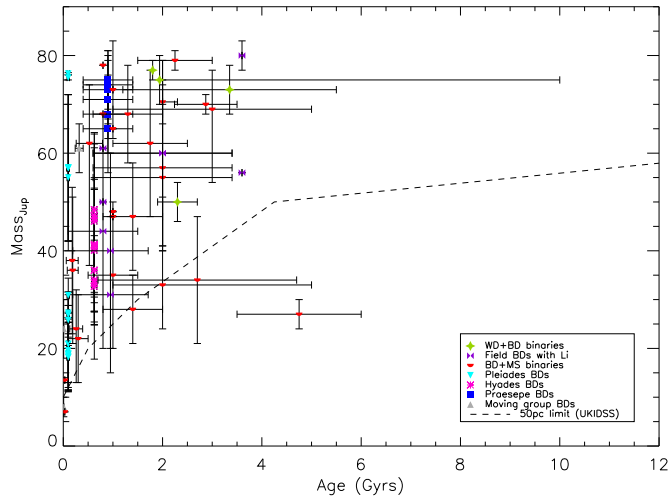}
\vspace{1.8cm}
\caption{The mass-age distribution of age benchmark brown dwarfs. Values taken from the literature.}
\label{fig:1}       % Give a unique label
\end{figure}
%\begin{table}
%\caption{Please write your table caption here.}
%\begin{tabular}{lll}
%\hline\noalign{\smallskip}
%\end{tabular}
%\end{table}
%

\end{document}